\documentclass[12pt]{iopart}

\usepackage{graphics}
\begin{document}

\title[Phenomenology of a three-family model with 3-4-1 gauge symmetry]{Phenomenology of a three-family model with gauge symmetry $SU(3)_c\otimes SU(4)_L\otimes U(1)_X$}

\author{Stiven Villada and Luis A. S\'anchez}

\address{Escuela de F\'\i sica, Universidad Nacional de Colombia,
A.A. 3840, Medell\'\i n, Colombia}
\eads{svillad@unal.edu.co and lasanche@unal.edu.co}

\begin{abstract}
We study an extension of the gauge group $SU(3)_c\otimes SU(2)_L\otimes U(1)_Y$ of the standard model to the symmetry group $SU(3)_c\otimes SU(4)_L\otimes U(1)_X$ (3-4-1 for short). This extension provides an interesting attempt to answer the question of family replication in the sense that models for the electroweak interaction can be constructed so that anomaly cancellation is achieved by an interplay between generations, all of them under the condition that the number of families must be divisible by the number of colours of $SU(3)_c$. This method of anomaly cancellation requires a family of quarks transforming differently from the other two, thus leading to tree-level flavour changing neutral currents (FCNC) transmitted by the two extra neutral gauge bosons $Z^\prime$ and $Z^{\prime \prime}$ predicted by the model. In a version of the 3-4-1 extension, which does not contain particles with exotic electric charges, we study the fermion mass spectrum and some aspects of the phenomenology of the neutral gauge boson sector. In particular, we impose limits on the $Z-Z^\prime$ mixing angle and on the mass scale of the corresponding physical new neutral gauge boson $Z_2$, and establish a lower bound on the mass of the additional new neutral gauge boson $Z^{\prime \prime} \equiv Z_3$. For the analysis we use updated precision electroweak data at the Z-pole from the CERN LEP and SLAC Linear Collider, and atomic parity violation data. The mass scale of the additional new neutral gauge boson $Z_3$ is constrained by using updated experimental inputs from neutral meson mixing in the analysis of the sources of FCNC in the model. The data constrain the $Z-Z^\prime$ mixing angle to a very small value of ${\cal O}(10^{-3})$, and the lower bounds on $M_{Z_2}$ and on $M_{Z_3}$ are found to be of ${\cal O}(1\; \mathrm{TeV})$ and of ${\cal O}(7\; \mathrm{TeV})$, repectively.
\end{abstract}

\maketitle

\section{Introduction}
\label{intro}
The number of fermion families in nature is one of the most intriguing puzzles in modern particle physics. Two possible scenarios for its solution have been proposed which relate the number of generations to cancellation of chiral anomalies. In one of them anomalies constrain the number of generations provided their cancellation takes place in an $SU(3)_c\otimes SU(2)_L\otimes U(1)_Y$ theory that lives in a six-dimensional spacetime \cite{dobrescu}. In the other one the standard model (SM) is extended either to the gauge group $SU(3)_c\otimes SU(3)_L\otimes U(1)_Y$ (the 3-3-1 model) \cite{su3,3-3-1} or to the gauge symmetry $SU(3)_c\otimes SU(4)_L\otimes U(1)_Y$ \cite{systematic,spp,swz,sen,sgp,su41,su42}, with anomalies cancelling among the families (three-family models) and not family by family as in the SM. In the 3-3-1 extension this happens only if we have an equal number of left handed triplets and antitriplets (taking into account the colour degree of freedom). Correspondingly, an equal number of 4-plets and $4^*$-plets is required in the 3-4-1 extension. As a consequence, the number of fermion families $N_f$ must be divisible by the number of colours $N_c$ of $SU(3)_c$, being $N_f=N_c=3$ the simplest solution.

In this work we will be concerned with 3-4-1 three-family models which do not contain particles with exotic electric charges. The systematic analysis of the 3-4-1 gauge theory carried out in \cite{systematic,sgp} has shown that the restriction to fermion field representations without exotic electric charges allows only for eight different anomaly free models. Four of them are three-family models and can be classified according to the values of two coefficients $b$ and $c$ which appear in the most general expression for the electric charge generator in $SU(4)_L\otimes U(1)_X$ (see (\ref{carga}) below).

The allowed simultaneous values for these coefficients, under the condition of absence of exotic electric charges, are: $b= c = 1$ and $b = 1,\; c = -2$ \cite{systematic}. Two of the four three-family models belong to the $b= c = 1$ class and have been studied in \cite{spp,swz,sen}. The other two models belong to the $b = 1,\; c = -2$ class, and one of them has been partially analyzed in \cite{sgp}. The other one, the so-called Model F in \cite{systematic}, has not been yet analyzed in the literature and will be studied in this work. These two classes of models differ both in their gauge and scalar boson sectors. So their also differ in their phenomenological implications.

One additional motivation to study the 3-4-1 theory comes from the fact that it has been recognized as a natural scenario for the implementation of the little Higgs mechanism \cite{little1}. Even though we will not be concerned here with this alternative proposal to solve the so-called hierarchy problem, we notice that in the simplest little Higgs scenario the SM gauge group is enlarged to $SU(3)_L\otimes U(1)_X$. This model, however, lacks a quartic Higgs coupling which can be generated in a $SU(4)_L\otimes U(1)_X$ extension \cite{little1}. The complete anomaly-free fermion sector for these two little Higgs models has been studied in detail in \cite{little2}, with direct generalization to $SU(N)_L\otimes U(1)_X$ with $N>4$ (at present, however, there is not motivation to go beyond $N=4$). Conspicuously, in the little Higgs scenario both the 3-3-1 extension and the 3-4-1 one are three-family models in which all the exotic fermion fields have only ordinary electric charges, and the complete anomaly-free fermion content obtained in \cite{little2} exactly coincides with the one obtained in \cite{3-3-1} and \cite{systematic}. 

3-4-1 models containing exotic electric charges have been also considered in the literature \cite{su41,su42}. In this case, a particular embedding of the SM gauge group into $SU(3)_c\otimes SU(4)_L\otimes U(1)_X$ depends on the physical motivation of the model to be constructed. The model in \cite{su41}, for example, has been proposed with the goal of including right-handed neutrinos in the fermion spectrum from the start. The supersymmetric extension of the 3-4-1 theory has also been explored \cite{rodri}.

The enlargement of the electroweak symmetry to $SU(4)_L\otimes U(1)_X$ leads to the prediction of two extra neutral gauge bosons $Z^\prime$ and $Z^{\prime \prime}$ which, in general, mix up with the known $Z$ boson of the SM. In models without exotic electric charges this mixing can be constrained to occur between $Z$ and $Z^\prime$ only, which leaves $Z^{\prime \prime} \equiv Z_3$ as a heavy mass eigenstate \cite{spp,swz,sgp,little1}. The diagonalization of the $Z-Z^\prime$ mass matrix produces a light mass eigentate $Z_1$, which can be identified as the neutral gauge boson of the SM, and a heavy $Z_2$. After the breakdown of the 3-4-1 symmetry down to $SU(3)_c\otimes U(1)_{Q}$, and since we have one family of quarks transforming differently from the other two under the gauge group, one important difference between the aforementioned two classes of three-family 3-4-1 models appears: even thought in both classes of models the $Z_1$ current is flavour diagonal, in the $b=c=1$ class the new $Z_2$ gauge boson couples nondiagonally to ordinary quarks thus transmitting tree-level FCNC at low energies, while the couplings to $Z_3$ are flavour diagonal. In the $b=1, \; c=-2$ class, instead, it is the new $Z_3$ gauge boson the responsible for this effect because couples nondiagonally to ordinary quarks, while the $Z_2$ current remains flavour diagonal.

As already mentioned, in this paper we will study a 3-4-1 model which belong to the $b=1,\; c=-2$ class. By using precision electroweak data at the Z-pole and atomic parity violation data, and by implementing a more appropriate computational approach than the one used in previous works, we do a $\chi^2$ fit to low energy data in order to set bounds on the $Z-Z^\prime$ mixing angle and on the mass of the new $Z_2$ gauge boson. Moreover, for the first time in the context of the $b=1, \; c=-2$ class of models, we obtain a lower bound on the mass of the new neutral gauge boson $Z_3$ by using updated experimental data coming from neutral meson mixing. The numerical approach used also allow us to perform a comparison between the predictions of the 3-4-1 model considered here and the predictions of the SM for a set of 22 low energy electroweak observables, an issue that, to the best of our knowledge, has not been previously addressed. Since the two models in the $b=1,\; c=-2$ class share the same gauge boson sector, the results we derive here correct, update and complete the ones obtained in \cite{spp}. In particular we show that the previous limits on the $Z_2$ mass are incorrect.

This paper is organized as follows. In the next section we introduce the model we are interested in by describing its anomaly-free fermion content, the scalar sector, and the mass spectrum in the gauge boson and fermion sectors. In section \ref{sec:3} we study the charged and neutral currents, paying special attention to the mixing in the neutral current sector. In section \ref{sec:4} we use electroweak precision measurements at the Z-pole, atomic parity violation (APV) data and experimental results from FCNC, in order to constrain the $Z-Z^\prime$ mixing angle and the mass scale of the new neutral gauge boson $Z_2$, as well as the mass scale of the $Z_3$ gauge boson predicted by the model. In the last section we summarize our results and state our conclusions.

\section{The model and its mass spectrum}
\label{sec:2}

Model F in \cite{systematic} is based on the local gauge symmetry $SU(3)_c\otimes SU(4)_L\otimes U(1)_X$ which contains $SU(3)_c\otimes SU(2)_L\otimes U(1)_X$ as a subgroup, and belongs to the $b=1,\; c= -2$ class, where $b$ and $c$ are parameters appearing in the expression for the electric charge operator in $SU(4)_L\otimes U(1)_X$
\begin{equation}\label{carga}
Q=a T_{3L}+\frac{b}{\sqrt{3}}T_{8L}+ \frac{c}{\sqrt{6}}T_{15L}+ XI_4,
\end{equation}
where $T_{iL}=\lambda_{iL}/2$, being $\lambda_{iL}$ the Gell-Mann matrices for $SU(4)_L$ normalized as Tr$(\lambda_i\lambda_j)=2\delta_{ij}$, $I_4=Dg(1,1,1,1)$ is the diagonal $4\times 4$ unit matrix, and $a=1$ gives the usual isospin of the electroweak interactions.

Its anomaly-free fermion structure has been already discussed in \cite{systematic} and is given in Table \ref{tab:1}, where $i=1,2$ and $\alpha=1,2,3$ are generation indexes. The numbers inside brackets correspond to the $SU(3)_c, SU(4)_L$ and $U(1)_X$ quantum numbers, respectively. $U_i$ and $U_3$ are exotic quarks of electric charge $2/3$, $D_i$ and $D_3$ are exotic quarks of electric charge $-1/3$, while $E^-_\alpha$ and $N^0_\alpha$ are exotic leptons.

We assume the symmetry breaking chain
\begin{eqnarray}\nonumber
SU(3)_c\otimes SU(4)_L\otimes & U(1)_X & \\ \nonumber
& \stackrel{V^\prime}{\longrightarrow}
& SU(3)_c\otimes SU(3)_L\otimes U(1)_Z \\ \nonumber
& \stackrel{V}{\longrightarrow} & SU(3)_c\otimes SU(2)_L\otimes U(1)_Y \\ \label{break}
& \stackrel{v+v^\prime}{\longrightarrow} & SU(3)_c\otimes U(1)_Q,
\end{eqnarray}
and we impose the hierarchy $V\sim V^\prime >> v\sim v^\prime \simeq 174$~GeV. This task is done by the following four Higgs scalars with vacuum expectation values (VEV) aligned as

\begin{table}
\caption{Anomaly free fermion content.}
\begin{indented}
\label{tab:1}
\item[]\begin{tabular}{ccccc}
\br
$Q_{iL}=\left(\begin{array}{c}d_i\\u_i\\U_i\\D_i \end{array}\right)_L$ &
$d^c_{iL}$ & $u^c_{iL}$ & $U^c_{iL}$ & $D^{c}_{iL}$ \\
$[3,4^*,\frac{1}{6}]$ & $[3^*,1,{1\over 3}]$ & $[3^*,1,-{2\over 3}]$
& $[3^*,1,-{2\over 3}]$ & $[3^*,1,{1\over 3}]$ \\ \mr
$Q_{3L}=\left(\begin{array}{c}u_3\\d_3\\D_3\\U_3
\end{array}\right)_L$ &
$u^c_{3L}$ & $d^c_{3L}$ & $D^c_{3L}$ & $U^{c}_{3L}$ \\
$[3,4,\frac{1}{6}]$ & $[3^*,1,-{2\over 3}]$ & $[3^*,1,{1\over 3}]$ & $[3^*,1,{1\over 3}]$ & $[3^*,1,-{2\over 3}]$ \\ \mr
$L_{\alpha L}=\left(\begin{array}{c}\nu _{e\alpha }^{0} \\e_{\alpha }^{-} \\E_{\alpha }^{-} \\N_{\alpha }^{0} \end{array}\right)_L $ &
$e^+_{\alpha L}$ & $E^+_{\alpha L}$ & $$ & $$ \\
$[1,4,-{1\over 2}]$ & $[1,1,1]$ & $[1,1,1]$ & $$ & $$\\
\br
\end{tabular}
\end{indented}
\end{table}

\begin{eqnarray}\nonumber
\left\langle \phi _{1}^{T}\right\rangle & =\left\langle \left( \phi
_{1}^{0},\phi _{1}^{+},\phi _{1}^{\prime +},\phi _{1}^{\prime 0}\right)
\right\rangle =\left( v,0,0,0\right) \sim \left[ 1,4^{\ast },1/2\right] , \\ \nonumber
\left\langle \phi _{2}^{T}\right\rangle & =\left\langle \left( \phi
_{2}^{-},\phi _{2}^{0},\phi _{2}^{\prime 0},\phi _{2}^{\prime -}\right)
\right\rangle =\left( 0,v^{\prime},0,0\right) \sim \left[ 1,4^{\ast },-1/2
\right] , \\ \nonumber
\left\langle \phi _{3}^{T}\right\rangle & =\left\langle \left( \phi
_{3}^{-},\phi _{3}^{0},\phi _{3}^{\prime 0},\phi _{3}^{\prime -}\right)
\right\rangle =\left( 0,0,V,0\right) \sim \left[ 1,4^{\ast },-1/2\right], \\ \label{vev}
\left\langle \phi _{4}^{T}\right\rangle & =\left\langle \left( \phi
_{4}^{0},\phi _{4}^{+},\phi _{4}^{\prime +},\phi _{4}^{\prime 0}\right)
\right\rangle =\left( 0,0,0,V^{\prime}\right) \sim \left[ 1,4^{\ast },1/2
\right],
\end{eqnarray}

We will see in what follows that this scalar structure provides masses for the gauge bosons and that, combined with a discrete symmetry, it is enough to produce a consistent mass spectrum for the charged fermion sector (quarks and leptons).

In this model there are a total of 24 gauge bosons which include one gauge field $B_{\mu}$ associated with $U(1)_{X}$, the 8 gluon fields associated with $SU(3)_{c}$, and 15 gauge fields associated with $SU(4)_{L}$. For $b=1$ and $c= -2$, the latter can be written for convenience as \cite{systematic}

\begin{equation}\label{gabo}
\frac{1}{2}\lambda _{L\alpha }A_{\mu }^{\alpha }=\frac{1}{\sqrt{2}}\left(
\begin{array}{cccc}
D_{1\mu }^{0} & W_{\mu }^{+} & K_{\mu }^{+} & X_{\mu }^{0} \\
W_{\mu }^{-} & D_{2\mu }^{0} & K_{\mu }^{0} & V_{\mu }^{-} \\
K_{\mu }^{-} & K_{\mu }^{\prime 0} & D_{3\mu }^{0} & Y_{\mu }^{-} \\
X_{\mu }^{\prime 0} & V_{\mu }^{+} & Y_{\mu }^{+} & D_{4\mu }^{0}
\end{array}
\right),
\end{equation}
where $D_{1}^{\mu }=A_{3}^{\mu }$ /$\sqrt{2}+A_{8}^{\mu }$ /$\sqrt{6}%
+A_{15}^{\mu }$ /$\sqrt{12}$, $D_{2}^{\mu }=-A_{3}^{\mu }$ /$\sqrt{2}%
+A_{8}^{\mu }$ /$\sqrt{6}+A_{15}^{\mu }$ /$\sqrt{12}$, $D_{3}^{\mu
}=-2A_{8}^{\mu }$ /$\sqrt{6}+A_{15}^{\mu }$ /$\sqrt{12}$ and $D_{4}^{\mu
}=-3A_{15}^{\mu }$ /$\sqrt{12}$.
\noindent
The covariant derivative for 4-plets is given by
\begin{equation}\label{covdev}
iD^{\mu }=i\partial ^{\mu }-g_{4}\lambda _{L\alpha }A_{\alpha }^{\mu}/2-g_{X}XB^{\mu }.
\end{equation}
\noindent
where $g_4$ and $g_X$ are the gauge coupling constants of the groups $SU(4)_{L}$ and $U(1)_X$, respectively. They obey the gauge matching conditions
\begin{equation}
g_{4}=g,\quad{\mbox{and}}\quad \frac{1}{g^{\prime 2}}=\frac{1}{g^{2}}+\frac{1}{%
g_{X}^{2}}, \\ \label{gmc}
\end{equation}
where $g$ and $g\prime$ are the gauge coupling constants of the $SU(2)_{L}$ and $U(1)_{Y}$ gauge groups of the SM, respectively.

A straightforward calculation shows that, after the 3-4-1 symmetry is broken with $\langle\phi_i\rangle, i=1,...,4$ and using the covariant derivative for $4$-plets given in (\ref{covdev}), the gauge boson $W^{\pm}$ does not mix with the other charged bosons and acquires a squared mass $M_{W^{\pm}}^{2}=(g_{4}^{2}/2)(v^{2}+v^{\prime 2})$. All the remaining charged bosons in the off-diagonal entries in (\ref{gabo}), namely: $K^{\pm}$, $V^{\pm}$, $Y^{\pm}$, $X^{0}(X^{\prime 0})$, and $K^{0}(K^{\prime 0})$, acquire masses at the large scale $V \sim V^\prime$. We can then identify $W^{\pm}$ as the charged gauge boson of the SM. Hence, with $g_4=g$ and using the experimental value $M_W=80.428\pm0.039$ GeV \cite{pdg}, we get $\sqrt{v^{2}+v^{\prime 2}}\approx v_{\mathrm{EW}}=174$ GeV.

In the neural gauge bosons sector, the massless photon $A^\mu$ and the massive bosons $Z^{\mu}$, $Z^{\prime\mu}$ and $Z^{\prime\prime\mu}$, are linear combinations of the diagonal entries in (\ref{gabo}) (for details see \cite{spp}). For the massive fields we have a $3\times3$ mass matrix in the basis $(Z^{\mu},Z^{\prime\mu},Z^{\prime\prime\mu})$. In the case we are considering, that is $V\simeq V^{\prime}$ and $v^{\prime}\simeq v$, the mixing between these three neutral gauge bosons simplifies. In fact, for this particular case the field $Z^{\prime \prime \mu}=\sqrt{2/3}A_{8}^{\mu}+A_{15}^{\mu}/{\sqrt{3}}\equiv Z_{3}^{\mu}$ does not mix with the other two and acquires a squared mass $M_{Z_{3}}^{2}=(g_{4}^{2}/2)(V^{2}+v^{2})$. This fact produces an enormous simplification in the study of the low energy deviations of the $Z$ couplings to the SM families which come from the diagonalization of the mass matrix  
\begin{equation}\label{remaining}
M_{(Z,Z^\prime)} = \frac{g_{4}^{2}}{C_{W}^{2}}\left(
\begin{array}{cc}
v^{2} & \delta v^{2}S_{W} \\
\delta v^{2}S_{W} & \qquad \frac{\delta ^{2}}{S_{W}^{2}}\left(
V^{2}C_{W}^{4}+v^{2}S_{W}^{4}\right)
\end{array}
\right),
\end{equation}
where $\delta =g_{X}/g_{4}$, and $S_{W}=\delta /\sqrt{2\delta ^{2}+1}$ and $C_{W}$ are the sine and cosine of the electroweak mixing angle, respectively. The corresponding mass eigenstates are: 
$Z_{1}^{\mu } = Z^{\mu }\cos \theta +Z^{\prime \mu}\sin \theta$ and 
$Z_{2}^{\mu } =-Z^{\mu }\sin \theta +Z^{\prime \mu}\cos \theta$,
where the mixing angle $\theta$ between $Z$ and $Z^{\prime}$ is given by
\begin{equation}
\tan (2\theta )=\frac{2S_{W}^{2}\sqrt{C_{2W}}}{2-(1+S_{W}^{2})^{2}+\frac{V^{2}
}{v^{2}}C_{W}^{4}}, \label{mix}
\end{equation}
with $C_{2W}=C_{W}^{2}-S_{W}^{2}$.

Concerning the fermion masses, in order to reduce the sources of FCNC in the model, we avoid mixing between ordinary and exotic fermions (which in turn avoids violation of the unitarity of the Cabibbo-Kobayashi-Maskawa (CKM) mixing matrix) by introducing an anomaly-free discrete $Z_2$ symmetry \cite{z2}. We assign $Z_2$ charges $q_{\mathrm{Z}}$ to the fields in the model as
\begin{eqnarray}
\nonumber
q_{\mathrm{Z}}(Q_{\alpha L},u_{\alpha L}^{c},d_{\alpha L}^{c},L_{\alpha
L},e_{\alpha L}^{c},\phi _{1},\phi _{2})&=& 0, \\ \label{Zcharge}
q_{\mathrm{Z}}(U_{\alpha L},D_{\alpha L}^{c},E_{\alpha L}^{c},\phi _{3},\phi
_{4})&=& 1,
\end{eqnarray}
where $\alpha=1,2,3$ is a family index as above.

It is easy to verify that the gauge invariance and the $Z_2$ symmetry do not allow for Yukawa terms in the neutral fermion sector. Hence, the neutral leptons in Table \ref{tab:1} remain massless. Notwithstanding, their masses and mixing can be implemented by introducing Weyl singlets with zero $X$-charges: $N_{L,n}^{0}\sim [1,1,0],\; n =1,2,...$, without violating the anomaly constraint relations. The appropriate implementation of masses and mixings can also require the enlargement of the scalar sector by including, for example, Higgs scalars belonging to the symmetric representation 10 of $SU(4)$, as shown in \cite{riazu}. 

For the charged leptons we find the following Yukawa terms
\begin{equation}\label{laglepton}
\bigskip \mathcal{L}_{Y}^{L}=\sum_{\alpha=1}^3\sum_{\beta=1}^3L_{\beta L}^{T}C[\phi _{2}h_{\alpha \beta }^{e}e_{\beta L}^{+}+\phi
_{3}h_{\alpha \beta }^{E}E_{\beta L}^{+}]+h.c.,
\end{equation}
where again the $h^\prime s$ are Yukawa couplings. From this equation we find a block diagonal mass matrix in the basis $(e_{1},e_{2},e_{3},E_{1},E_{2},E_{3})$, given by
\begin{equation}\label{masslepton}
M_{eE}=\left(\begin{array}{cc}
M_{3\times 3}^{e} & 0 \\
0 & M_{3\times 3}^{E}%
\end{array}%
\right),
\end{equation}
where the entries in the submatrices are
\begin{equation}\label{subm}
M^e_{\alpha \beta }=h_{\alpha \beta }^{e}v^{\prime} \quad{\mbox{and}}\quad M^E_{\alpha \beta
}=h_{\alpha \beta }^{E}V.
\end{equation}
  
For the quark sector, we identify the following Yukawa terms
\begin{eqnarray}\nonumber
\bigskip \mathcal{L}_{Y}^{Q} &=&\sum_{i=1}^2Q_{iL}^{T}C[\phi
_{2}^{\ast }\sum_{\alpha=1}^3h_{i\alpha }^{u}u_{\alpha
L}^{c}+\phi _{1}^{\ast }\sum_{\alpha=1}^3h_{i\alpha
}^{d}d_{\alpha L}^{c}+\phi _{3}^{\ast }\sum_{\alpha=1}^3h_{i\alpha }^{U}U_{\alpha L}^{c} \\ \nonumber
&&+\phi _{4}^{\ast }\sum_{\alpha=1}^3h_{i\alpha }^{D}D_{\alpha
L}^{c}]+Q_{3L}^{T}C[\phi _{1}\sum_{\alpha=1}^3h_{i\alpha
}^{u}u_{\alpha L}^{c}+\phi _{2}\sum_{\alpha=1}^3h_{3\alpha
}^{d}d_{\alpha L}^{c} \\ \label{lagquark}
&&+\phi _{4}\sum_{\alpha=1}^3h_{3\alpha }^{U}U_{\alpha
L}^{c}+\phi _{3}\sum_{\alpha=1}^3h_{3\alpha }^{D}D_{\alpha
L}^{c}]+h.c.,
\end{eqnarray}
where the $h^\prime s$ are Yukawa couplings and $C$ is the charge conjugation operator. From this Lagragian we get, for the up- and down-type quarks in the basis $(u_{1},u_{2},u_{3},U_{1},U_{2},U_{3})$ and $(d_{1},d_{2},d_{3},D_{1},D_{2},D_{3})$, respectively, $6\times 6$ block diagonal mass matrices of the form

\begin{equation}\label{massquark}
M_{uU}=\left(\begin{array}{cc}M_{3\times 3}^{u} & 0 \\
0 & M_{3\times 3}^{U}
\end{array}
\right) \quad \mathrm{and} \quad M_{dD}=\left(\begin{array}{cc}M_{3\times 3}^{d} & 0 \\
0 & M_{3\times 3}^{D}
\end{array}
\right),
\end{equation}
where
\begin{equation}\label{mu}
M^{u}=\left(
\begin{array}{ccc}
h_{11}^{u}v^{\prime} & h_{12}^{u}v^{\prime} & h_{13}^{u}v^{\prime} \\
h_{21}^{u}v^{\prime} & h_{22}^{u}v^{\prime} & h_{23}^{u}v^{\prime} \\
h_{31}^{u}v & h_{32}^{u}v & h_{33}^{u}v
\end{array}
\right),\quad M^{U}=\left(
\begin{array}{ccc}
h_{11}^{U}V & h_{12}^{U}V & h_{13}^{U}V \\
h_{21}^{U}V & h_{22}^{U}V & h_{23}^{U}V \\
h_{31}^{U}V^{\prime} & h_{32}^{U}V^{\prime} & h_{33}^{U}V^{\prime}
\end{array}
\right)
\end{equation}
\begin{equation}\label{md}
M^{d}=\left(
\begin{array}{ccc}
h_{11}^{d}v & h_{12}^{d}v & h_{13}^{d}v \\
h_{21}^{d}v & h_{22}^{d}v & h_{23}^{d}v \\
h_{31}^{d}v^{\prime} & h_{32}^{d}v^{\prime} & h_{33}^{d}v^{\prime}
\end{array}
\right), \quad M^{D}=\left(
\begin{array}{ccc}
h_{11}^{D}V^{\prime} & h_{12}^{D}V^{\prime} & h_{13}^{D}V^{\prime} \\
h_{21}^{D}V^{\prime} & h_{22}^{D}V^{\prime} & h_{23}^{D}V^{\prime} \\
h_{31}^{D}V & h_{32}^{D}V & h_{33}^{D}V
\end{array}
\right) .
\end{equation}

These mass matrices show that all the charged fermions in the model acquire masses at the three level, and that all the ordinary fermions get masses at the low scale $v^\prime \simeq v$, while all the exotic fermions acquire masses at the high scale $V\sim V^\prime$. The unitarity of the CKM mixing matrix is guaranteed because the tensor product form of the mass matrices $M_{uU}$ and $M_{dD}$ in (\ref{massquark}) implies that they are diagonalized by unitary matrices which are themselves tensor products of unitary matrices.


\section{Currents}
\label{sec:3}
\subsection{Charged currents}
\label{sec:31}

The charged currents Lagrangian looks like
\begin{eqnarray*}
-\mathcal{L}^{\mathrm{CC}}&=&\frac{g_{4}}{\sqrt{2}}(W_{\mu }^{+}J_{W^{+}}^{\mu }+K_{\mu
}^{+}J_{K^{+}}^{\mu }+V_{\mu }^{+}J_{V^{+}}^{\mu }+Y_{\mu
}^{+}J_{W^{+}}^{\mu } \\
&&+X_{\mu }^{0}J_{X^{0}}^{\mu }+K_{\mu
}^{0}J_{K^{0}}^{\mu })+h.c.,
\end{eqnarray*}
where the currents are
\begin{eqnarray}\nonumber
J_{W^{+}}^{\mu } &=&\bar{u}_{3L}\gamma ^{\mu }d_{3L}-\sum_{i=1}^2\bar{u}
_{iL}\gamma ^{\mu }d_{iL}+\sum_{\alpha=1}^3\bar{\nu }_{\alpha L}\gamma ^{\mu }e_{\alpha
L}^{-}, \\ \nonumber
J_{K^{+}}^{\mu } &=&\bar{u}_{3L}\gamma ^{\mu }D_{3L}-\sum_{i=1}^2\bar{U}
_{iL}\gamma ^{\mu }d_{iL}+\sum_{\alpha=1}^3\bar{\nu }_{\alpha L}\gamma ^{\mu }E_{\alpha
L}^{-}, \\ \nonumber
J_{V^{+}}^{\mu } &=&\bar{u}_{3L}\gamma ^{\mu }d_{3L}-\sum_{i=1}^2\bar{u}
_{iL}\gamma ^{\mu }D_{iL}+\sum_{\alpha=1}^3\bar{N}_{\alpha L}^{0}\gamma ^{\mu }e_{\alpha
L}^{-}, \\ \nonumber
J_{Y^{+}}^{\mu } &=&\bar{u}_{3L}\gamma ^{\mu }D_{3L}-\sum_{i=1}^2\bar{U}
_{iL}\gamma ^{\mu }D_{iL}+\sum_{\alpha=1}^3\bar{N}_{\alpha L}^{0}\gamma ^{\mu }E_{\alpha
L}^{-}, \\ \nonumber
J_{X^{0}}^{\mu } &=&\bar{u}_{3L}\gamma ^{\mu }U_{3L}-\sum_{i=1}^2\bar{D}
_{iL}\gamma ^{\mu }d_{iL}+\sum_{\alpha=1}^3\bar{\nu }_{\alpha L}\gamma ^{\mu }N_{\alpha
L}^{0}, \\ \label{chcurr}
J_{K^{0}}^{\mu } &=&\bar{d}_{3L}\gamma ^{\mu }D_{3L}-\sum_{i=1}^2\bar{U}
_{iL}\gamma ^{\mu }u_{iL}+\sum_{\alpha=1}^3\bar{e}_{\alpha L}^{-}\gamma ^{\mu }E_{\alpha
L}^{-}.
\end{eqnarray}


\subsection{Neutral currents}
\label{sec:32}

The Lagrangian for the neutral currents $J_{\mu}(\mathrm{EM})$, $J_{\mu}(Z)$, 
$J_{\mu}(Z^{\prime})$, and $J_{\mu}(Z^{\prime \prime})$ is written as
\begin{eqnarray}\nonumber
-\mathcal{L}^{\mathrm{NC}}&=&eA^{\mu }J_{\mu }(\mathrm{EM})+(g_{4}/C_{W})Z^{\mu}J_{\mu}(Z)+g_{X}Z^{\prime\mu}J_{\mu}(Z^{\prime}) \\ \label{LNC}
&&+g_{4}/(2\sqrt{2})Z^{\prime \prime}J_{\mu}(Z^{\prime \prime}),
\end{eqnarray}
with
\begin{eqnarray}\nonumber
J_{\mu}(\mathrm{EM}) &=&\frac{2}{3}[\bar{u}_{3}\gamma _{\mu}u_{3}+\bar{U}
_{3}\gamma _{\mu}U_{3}+\sum_{i=1}^2(\bar{u}_{i}\gamma _{\mu}u_{i}+\bar{U}
_{i}\gamma _{\mu}U_{i})] \\ \nonumber
&&-\frac{1}{3}[\bar{d}_{3}\gamma _{\mu }d_{3}+\bar{D}_{3}\gamma
_{\mu }D_{3}+\sum_{i=1}^2(\bar{d}_{i}\gamma _{\mu }d_{i}+\bar{D}_{i}\gamma
_{\mu }D_{i})] \\ \nonumber
&&-\sum_{\alpha=1}^3(\bar{e}_{\alpha}^{-}\gamma _{\mu }e_{\alpha }^{-}+\bar{E}
_{\alpha }^{-}\gamma _{\mu }E_{\alpha}^{-}) \\ \label{emcurr}
&=& \sum_{f} q_{f}\bar{f}\gamma _{\mu }f,
\end{eqnarray}
\begin{equation}\label{zeta}
J_{\mu}(Z)=J_{\mu ,L}(Z)-S_{W}^{2}J_{\mu }(\mathrm{EM}),
\end{equation}
\begin{equation}\label{zetap}
J_{\mu}(Z^{\prime})=J_{\mu,L}(Z^{\prime})-T_{W}J_{\mu}(\mathrm{EM}),
\end{equation}
\begin{eqnarray}\nonumber
J_{\mu}(Z^{\prime \prime}) &=&\sum_{i=1}^2(-\bar{d}_{iL}\gamma _{\mu}d_{iL}-
\bar{u}_{iL}\gamma _{\mu }u_{iL}+\bar{U}_{iL}\gamma _{\mu }U_{iL}+\bar{D}_{iL}\gamma _{\mu }D_{iL}) \\ \nonumber
&&+\bar{u}_{3L}\gamma _{\mu
}u_{3L}+\bar{d}_{3L}\gamma _{\mu }d_{3L}-\bar{D}_{3L}\gamma _{\mu }D_{3L}-\bar{U}_{3L}\gamma _{\mu}U_{3L} \\ \label{zprimaprima}
&&+\sum_{\alpha=1}^3(\bar{\nu }_{\alpha L}\gamma _{\mu }\nu _{\alpha L}+\bar{e}
_{\alpha L}^{-}\gamma _{\mu }e_{\alpha L}^{-}-\bar{E}_{\alpha L}^{-}\gamma _{\mu }E_{\alpha L}^{-}-\bar{N}
_{\alpha L}^{0}\gamma _{\mu }N_{\alpha L}^{0}), 
\end{eqnarray}
where $e=gS_{W}=g_{X}C_{W}\sqrt{1-T_{W}^{2}}>0$, and $q_f$ is the electric charge of the fermion $f$ in units of $e$. Note that $J_{\mu}(Z^{\prime \prime})$ is a pure left-handed current and that, notwithstanding the neutral gauge boson $Z_{\mu}^{\prime \prime}$ does not mix neither with $Z_{\mu}$ nor with $Z_{\mu}^{\prime}$ (for the particular case $V\simeq V^\prime$ and $v\simeq v^\prime$), it still couples nondiagonally to ordinary fermions. As a matter of fact, its couplings to the third family of quarks are different from the ones to the first two families. Thus, at low energy, we have tree-level FCNC transmitted by $Z_{\mu }^{\prime \prime}$. This is in contrast with the $b=c=1$ class of 3-4-1 three-family models where the tree-level FCNC are transmitted by the $Z_{\mu }^{\prime}$ gauge boson.

The two neutral left-handed currents in $J_{\mu}(Z)$ and $J_{\mu}(Z^{\prime})$ are given by
\begin{eqnarray}\nonumber
J_{\mu ,L}(Z) &=&\frac{1}{2}[\bar{u}_{3L}\gamma _{\mu }u_{3L}-\bar{
d}_{3L}\gamma _{\mu }d_{3L}-\sum_{i=1}^2(\bar{d}_{iL}\gamma _{\mu }d_{iL}-\bar{u}_{iL}\gamma _{\mu}u_{iL}) \\ \label{zleft}
&&+\sum_{\alpha=1}^3(\bar{\nu }_{\alpha L}\gamma _{\mu }\nu _{\alpha L}-\bar{e}
_{\alpha L}^{-}\gamma _{\mu }e_{\alpha L}^{-})], 
\end{eqnarray}
\begin{eqnarray}\nonumber
J_{\mu ,L}(Z^{\prime }) &=&(2T_{W})^{-1}[T_{W}^{2}\bar{u}_{3L}\gamma
_{\mu }u_{3L}-T_{W}^{2}\bar{d}_{3L}\gamma _{\mu }d_{3L}
-\bar{D}_{3L}\gamma _{\mu }D_{3L} \\ \nonumber
&&+\bar{U}_{3L}\gamma _{\mu}U_{3L}-\sum_{i=1}^2(T_{W}^{2}\bar{d}_{iL}\gamma _{\mu }d_{iL}-T_{W}^{2}\bar{u}
_{iL}\gamma _{\mu }u_{iL} \\ \nonumber
&&-\bar{U}_{iL}\gamma _{\mu }U_{iL}+\bar{D}_{iL}\gamma _{\mu
}D_{iL})+\sum_{\alpha=1}^3(T_{W}^{2}\bar{\nu }_{\alpha L}\gamma _{\mu }\nu _{\alpha
L} \\ \label{zprimaleft}
&&-T_{W}^{2}\bar{e}_{\alpha L}^{-}\gamma _{\mu }e_{\alpha L}^{-}-\bar{E}_{\alpha L}^{-}\gamma _{\mu }E_{\alpha L}^{-}+\bar{N}
_{\alpha L}^{0}\gamma _{\mu }N_{\alpha L}^{0})]. 
\end{eqnarray}
Since $J_{\mu}(Z)$ is the generalization of the neutral current of the SM, we can identify $Z_\mu$ as the neutral gauge boson of the SM. From (\ref{zprimaleft}) we realize that the neutral gauge boson $Z_\mu^\prime$ does not transmit FCNC at low energy because couples diagonally to ordinary fermions.

The couplings between the fermion fields and the mass eigenstates $Z_1^{\mu}$, $Z_2^{\mu}$ are extracted from the second and third terms in (\ref{LNC}) written in the $V-A$ form
\begin{equation} \nonumber
-\mathcal{L}_{Z_{1},Z_{2}}^{\mathrm{NC}}=\frac{g_{4}}{2C_{W}}\sum_{i=1}^2Z_{i}^{\mu }\sum_{f}\left\{ \bar{f}\gamma^{\mu }\left[ g(f)_{iV}-g(f)_{iA}\gamma ^{5}\right] f\right\} ,
\end{equation}
where
\begin{eqnarray}\nonumber
g(f)_{1V} &=&\cos \theta \left( T_{4f}-2q_{f}S_{W}^{2}\right) +\frac{g_{X}}{
g_{4}}\sin \theta \left( T_{4f}^{\prime }C_{W}-2q_{f}S_{W}\right), \\ \nonumber
g(f)_{1A} &=&\cos \theta T_{4f}+\frac{g_{X}}{g_{4}}\sin \theta
T_{4f}^{\prime }C_{W}, \\ \nonumber
g(f)_{2V} &=&-\sin \theta \left( T_{4f}-2q_{f}S_{W}^{2}\right) +\frac{g_{X}}{
g_{4}}\cos \theta \left( T_{4f}^{\prime }C_{W}-2q_{f}S_{W}\right), \\ \label{coup}
g(f)_{2A} &=&-\sin \theta T_{4f}+\frac{g_{X}}{g_{4}}\cos \theta
T_{4f}^{\prime }C_{W}.
\end{eqnarray}
Here, $T_{4f}=Dg(1/2,-1/2,0,0)$ is the third component of the weak isospin and $T_{4f}^{\prime}=(1/2T_{W})Dg(T_{W}^{2},-T_{W}^{2},-1,1)=T_{W}\lambda_{3}/2+(1/T_{W})(\lambda _{8}/(2\sqrt{3})-\lambda _{15}/\sqrt{6})$. The expresions for $g(f)_{iV},\; g(f)_{iA}$, $i=1,2$ for all the fermions in the model are listed in Tables \ref{tab:2} and \ref{tab:3}, where $\Upsilon =1/\sqrt{1-2S_{W}^{2}}$. From these Tables we see that the couplings are family-universal. This is a direct consequence of the fact that, according Table~\ref{tab:1}, the three families of quarks have the same hypercharge $X$.

Note that in the limit $\theta \rightarrow 0$ the couplings of $Z_1^{\mu}$ to ordinary quarks and leptons are the same that in the SM. This will allows us, in the next section, to test the new physics predicted by the 3-4-1 extension we are studying.

\begin{table}
\caption{The $Z_1^\mu\longrightarrow \bar{f}f$ couplings.}
\begin{indented}
\label{tab:2}
\item[]\begin{tabular}{lcl}
\br
$f$ & $g(f)_{1V}$ & $g(f)_{1A}$ \\
\mr
$u_{1,2,3}$ & $\left( \frac{1}{2}-\frac{4S_{W}^{2}}{3}\right) \cos \theta -\frac{
5\Upsilon S_{W}^{2}}{6}\sin \theta $ & $\frac{1}{2}\left( \cos \theta
+\Upsilon S_{W}^{2}\sin \theta \right) $ \\
$d_{1,2,3}$ & $\left( -\frac{1}{2}+\frac{2S_{W}^{2}}{3}\right) \cos \theta +
\frac{\Upsilon S_{W}^{2}}{6}\sin \theta $ & $-\frac{1}{2}\left( \cos \theta
+\Upsilon S_{W}^{2}\sin \theta \right) $ \\
$D_{1,2,3}$ & $\frac{2S_{W}^{2}}{3}\cos \theta +\frac{\Upsilon }{6}\left(
-3C_{W}^{2}+4S_{W}^{2}\right) \sin \theta $ & $-\frac{1}{2}\Upsilon
C_{W}^{2}\sin \theta $ \\
$U_{1,2,3}$ & $ -\frac{4S_{W}^{2}}{3} \cos \theta +\frac{
\Upsilon }{6}\left( 3C_{W}^{2}-8S_{W}^{2}\right) \sin \theta $ & $
\frac{1}{2}\Upsilon C_{W}^{2}\sin \theta $ \\
$\nu _{1,2,3}$ & $\frac{1}{2}\cos \theta +\frac{\Upsilon }{2}S_{W}^{2}\sin
\theta $ & $\frac{1}{2}\left( \cos \theta +\Upsilon S_{W}^{2}\sin \theta
\right) $ \\
$e_{1,2,3}^{-}$ & $\left( -\frac{1}{2}+2S_{W}^{2}\right) \cos \theta +\frac{
3\Upsilon }{2}S_{W}^{2}\sin \theta $ & $-\frac{1}{2}\left( \cos \theta
+\Upsilon S_{W}^{2}\sin \theta \right) $ \\
$E_{1,2,3}^{-}$ & $2S_{W}^{2}\cos \theta +\frac{\Upsilon }{2}\left(
-C_{W}^{2}+4S_{W}^{2}\right) \sin \theta $ & $-\frac{1}{2}\Upsilon
C_{W}^{2}\sin \theta $ \\
$N_{1,2,3}$ & $\frac{\Upsilon }{2}C_{W}^{2}\sin \theta $ & $\frac{1}{2}
\Upsilon C_{W}^{2}\sin \theta $ \\
\br
\end{tabular}
\end{indented}
\end{table}

\begin{table}
\caption{The $Z_2^\mu\longrightarrow \bar{f}f$ couplings.}
\begin{indented}
\label{tab:3}
\item[]\begin{tabular}{lcl}
\br
$f$ & $g(f)_{2V}$ & $g(f)_{2A}$ \\
\mr
$u_{1,2,3}$ & $-\left( \frac{1}{2}-\frac{4S_{W}^{2}}{3}\right) \sin \theta -
\frac{5\Upsilon S_{W}^{2}}{6}\cos \theta $ & $\frac{1}{2}\left( -\sin \theta
+\Upsilon S_{W}^{2}\cos \theta \right) $ \\
$d_{1,2,3}$ & $-\left( -\frac{1}{2}+\frac{2S_{W}^{2}}{3}\right) \sin \theta +
\frac{\Upsilon S_{W}^{2}}{6}\cos \theta $ & $-\frac{1}{2}\left( -\sin \theta
+\Upsilon S_{W}^{2}\cos \theta \right) $ \\
$D_{1,2,3}$ & $-\frac{2S_{W}^{2}}{3}\sin \theta +\frac{\Upsilon }{6}\left(
-3C_{W}^{2}+4S_{W}^{2}\right) \cos \theta $ & $-\frac{1}{2}\Upsilon
C_{W}^{2}\cos \theta $ \\
$U_{1,2,3}$ & $\frac{4S_{W}^{2}}{3}\sin \theta +\frac{\Upsilon }{6}\left(
3C_{W}^{2}-8S_{W}^{2}\right) \cos \theta $ & $\frac{1}{2}\Upsilon
C_{W}^{2}\cos \theta $ \\
$\nu _{1,2,3}$ & $-\frac{1}{2}\sin \theta +\frac{\Upsilon }{2}S_{W}^{2}\cos
\theta $ & $\frac{1}{2}\left( -\sin \theta +\Upsilon S_{W}^{2}\cos \theta
\right) $ \\
$e_{1,2,3}^{-}$ & $-\left( -\frac{1}{2}+2S_{W}^{2}\right) \sin \theta +\frac{
3\Upsilon }{2}S_{W}^{2}\cos \theta $ & $-\frac{1}{2}\left( -\sin \theta
+\Upsilon S_{W}^{2}\cos \theta \right) $ \\
$E_{1,2,3}^{-}$ & $-2S_{W}^{2}\sin \theta +\frac{\Upsilon }{2}\left(
-C_{W}^{2}+4S_{W}^{2}\right) \cos \theta $ & $-\frac{1}{2}\Upsilon
C_{W}^{2}\cos \theta $ \\
$N_{1,2,3}$ & $\frac{\Upsilon }{2}C_{W}^{2}\cos \theta $ & $\frac{1}{2}
\Upsilon C_{W}^{2}\cos \theta $ \\
\br
\end{tabular}
\end{indented}
\end{table}

\section{Low energy constraints on the parameters of the model}
\label{sec:4}

\subsection{Bounds on $M_{Z_2}$ and $\theta$ from Z-pole observables and APV data}
\label{sec:41}

To get bounds on the parameter space $(\theta-M_{Z_2})$ and to test the model by low energy data, we use electroweak observables measured at the $Z$-pole from the CERN $e^+e^-$ collider (LEP), SLAC Linear Collider (SLC), and atomic parity violation data which are given in Table \ref{tab:4} \cite{pdg}. Let us start by briefly describing each one of the observables in the Table.

The expression for the partial decay width for the gauge boson $Z_1^\mu$ to decay into massless ordinary SM fermions $f\bar{f}$, including the electroweak and QCD virtual corrections is given, in the on-shell scheme, by \cite{pdg,bernabeu}
\begin{eqnarray}\nonumber
\Gamma(Z^{\mu}_1\rightarrow f\bar{f})&=&\frac{N_C G_F
M_{Z_1}^3}{6\pi\sqrt{2}}\rho_f \Big\{\frac{3\beta-\beta^3}{2}
[g(f)_{1V}]^2 \\ \label{decz}
& + & \beta^3[g(f)_{1A}]^2 \Big\}(1+\delta_f)R_{\mathrm{EW}}R_{\mathrm{QCD}}.
\end{eqnarray}
In the modified minimal substraction $(\overline{\mbox{MS}})$ scheme, which we use through this section, the normalization is changed according to $G_F M_{Z_1}^2/(2\sqrt{2}\pi) \rightarrow \widehat{\alpha}/\lbrack 4 \sin^2\widehat{\theta}_W(M_{Z_1}) \cos^2\widehat{\theta}_W(M_{Z_1})\rbrack$. In (\ref{decz}), $Z^\mu_1$ is the physical gauge boson observed at LEP, $N_C=1$ for leptons while for quarks $N_C=3(1+\alpha_s/\pi + 1.405\alpha_s^2/\pi^2 - 12.77\alpha_s^3/\pi^3)$,
where the $3$ is due to colour and the factor in parentheses represents the
universal part of the QCD corrections for massless quarks. $R_{\mathrm{EW}}$ are electroweak corrections which include the leading order QED corrections given by $R_{\mathrm{QED}}$ $=1+3\alpha q^2_f/(4\pi)$. $R_{\mathrm{QCD}}$ are further QCD corrections, and $\beta=\sqrt{1-4 m_f^2/M_{Z_1}^2}$ is a kinematic factor which can be taken equal to $1$ for all the SM fermions except for the bottom quark. The parameter $\rho_f$ is written as $\rho_f = 1+\rho_t$ where $\rho_t = 3G_F m_t^2/(8\pi^2\sqrt{2})$ with $m_t$ being the top quark pole mass. Universal electroweak corrections are included in $\rho_t$, and in the coupling constants $g(f)_{1V}$ and $g(f)_{1A}$ of the physical $Z_1^\mu$ field with ordinary fermions which are written in terms of the effective electroweak mixing angle $\bar{S}^2_W = \kappa_f S^2_W \approx (1+\rho_t/T^2_W)S^2_W$. In the $\overline{\mbox{MS}}$ scheme, $\widehat{\rho}_f \sim 1$ and $\widehat{\kappa}_f \sim 1$ for $f\neq b$, while $\widehat{\rho}_b \sim 1 - (4/3)\rho_t$ and $\widehat{\kappa}_b \sim 1 + (2/3)\rho_t$. The factor $\delta_f$ contains the one loop vertex contribution which is negligible for all fermion fields except for the bottom quark for which the contribution coming from the top quark at the one loop vertex radiative correction is parametrized as $\delta_b\approx 10^{-2} [-m_t^2/(2 M_{Z_1}^2)+1/5]$. In the $\overline{\mbox{MS}}$ scheme this correction is included in $\widehat{\rho}_b$ and $\widehat{\kappa}_b$.

The total hadronic cross-section is
\begin{equation}\label{cross}
\sigma_{\mathrm{had}}=\frac{12\pi}{M^2_{Z_1}}\frac{\Gamma_{(e^+e^-)}\Gamma(\mathrm{had})}{\Gamma^2_Z},
\end{equation}
where $\Gamma_Z$ is the total width for $Z^{\mu}_1\rightarrow f\bar{f}$.

The ratios of partial widths are defined as
\begin{equation}\label{ratio1}
R_l\equiv \frac{\Gamma(\mathrm{had})}{\Gamma(l^+l^-)}\quad \mbox{for}\quad l=e,\mu,\tau,
\end{equation}
and
\begin{equation}\label{ratio2}
R_\eta\equiv \frac{\Gamma_\eta}{\Gamma(\mathrm{had})}\quad \mbox{for}\quad \eta=b,c.
\end{equation}

The forward-backward asymmetries at the Z-pole are given by
\begin{eqnarray}\label{fbasymm}
A^{(0,f)}_{FB}=\frac{3}{4}A_eA_f,& \quad{\mbox{where}}\quad &A_f=\frac{2g(f)_{1V}g(f)_{1A}}{g(f)^{2}_{1V}+g(f)^{2}_{1A}}
\end{eqnarray}
($f= e, \mu, \tau, s, c, b$), which are also written in terms of $\bar{S}^2_W$.

\begin{table}
\caption{Experimental data and SM values for the observables used for the $\chi^2$ fit.}
\begin{indented}
\label{tab:4}
\item[]\begin{tabular}{lcc}
\br
& Experimental results & SM value\\
\mr
$\Gamma _{Z}$\; [\mbox{GeV}] & $2.4952\pm 0.0023$ & $2.4968\pm 0.0010$ \\
$\Gamma (\mbox{had})$\; [\mbox{GeV}] & $1.7444\pm 0.0020$ & $1.7434\pm 0.0010$ \\
$\Gamma_{(l^{+}l^{-})}\; [\mbox{MeV}]$ & $83.984\pm 0.086$ & $83.988\pm 0.016$ \\
$\sigma_{\mathrm{had}}\; \lbrack \mbox{nb} \rbrack$ & $41.541 \pm 0.037$ & $41.466 \pm 0.009$ \\
$R_e$ & $20.804\pm 0.050$ & $20.758\pm 0.011$ \\
$R_{\mu}$ & $20.785\pm 0.033$ & $20.758\pm 0.011$ \\
$R_{\tau}$ & $20.764\pm 0.045$ & $20.803\pm 0.011$ \\
$R_{b}$ & $0.21629\pm 0.00066$ & $0.21584\pm 0.00006$ \\
$R_{c}$ & $0.1721\pm 0.0030$ & $0.17228\pm 0.00004$ \\
$A_{FB}^{(0,e)}$ & $0.0145\pm 0.0025$ & $0.01627\pm 0.00023$ \\
$A_{FB}^{(0,\mu )}$ & $0.0169\pm 0.0013$ & \\
$A_{FB}^{(0,\tau )}$ & $0.0188\pm 0.0017$ & \\
$A_{FB}^{(0,b)}$ & $0.0992\pm 0.0016$ & $0.1033\pm 0.0007$ \\
$A_{FB}^{(0,c)}$ & $0.0707\pm 0.0035$ & $0.0738\pm 0.0006$ \\
$A_{FB}^{(0,s)}$ & $0.0976\pm 0.0114$ & $0.1034\pm 0.0007$ \\
$A_{e}$ & $0.15138\pm 0.00216$ & $0.1473\pm 0.0011$ \\
$A_{\mu}$ & $0.142\pm 0.015$ & \\
$A_{\tau}$ & $0.136\pm 0.015$ & \\
$A_{b}$ & $0.923\pm 0.020$ & $0.9347\pm 0.0001$ \\
$A_{c}$ & $0.670\pm 0.027$ & $0.6678\pm 0.0005$ \\
$A_{s}$ & $0.895\pm 0.091$ & $0.9536\pm 0.0001$ \\
$Q_{W}(\mathrm{Cs})$ & $-72.62\pm 0.46$ & $-73.16\pm 0.03$ \\
\br
\end{tabular}
\end{indented}
\end{table}

The 3-4-1 new physics effects on the SM observables listed in the first column of Table \ref{tab:5} are obtained by noticing that, with the assumed hierarchy $V>>v$ and from (\ref{mix}), the $Z-Z^\prime$ mixing angle is expected to be very small so, $\cos\theta =\sqrt{1-\sin^2\theta} \simeq 1-(1/2)\sin^2\theta \simeq 1$, and the coupling constants $g(f)_{1V}$ and $g(f)_{1A}$ of the physical $Z_1^\mu$ gauge boson to ordinary fermions can be written as (remember that in the limit $\theta \rightarrow 0$ these couplings are the same as in the SM)
\begin{equation}\label{g331}
g(f)_{1V,A}= g(f)^{\mathrm{SM}}_{1V,A} + \delta g(f)_{1V,A},
\end{equation}
where the expressions for $\delta g(f)_{1V,A}$ depend lineary on $\sin\theta$ and can be easily read from Table \ref{tab:2}.

To facilitate the numerical analysis we express the changes in the physical observables relative to their SM values as \cite{Malkawi}

\begin{equation}\label{corr}
O^{341}= O^{\mathrm{SM}}(1 + \delta_O), \quad \mbox{where} \quad \delta_O = 
\frac{\delta O}{O^{\mathrm{SM}}},
\end{equation}
with $O^{\mathrm{SM}}$ being the SM value for the observable $O$, including the one-loop SM corrections, and with $\delta O$ representing the corrections due to new physics. Equation (\ref{corr}) allows us to quickly assess the percentage changes in the SM observables brought about by the various 3-4-1 corrections. 

For the observables in Table \ref{tab:4}, and taking into account from Table \ref{tab:2} that the couplings $g(f)_{1V}$ and $g(f)_{1A}$ are family universal, the several $\delta_O$ are given by

\begin{eqnarray}\label{corr1}
\delta_Z &=& \frac{1}{\Gamma^{\mathrm{SM}}_Z} \bigl(2 \Gamma^{\mathrm{SM}}_u \delta_u + 2\Gamma^{\mathrm{SM}}_d \delta_d  + \Gamma^{\mathrm{SM}}_b\delta_b  + 3\Gamma^{\mathrm{SM}}_\nu \delta_\nu  + 3\Gamma^{\mathrm{SM}}_e \delta_l \bigr), \\ \label{corr2}
\delta_{\mathrm{had}} &=& 2 R^{\mathrm{SM}}_c \delta_u + R^{\mathrm{SM}}_b \delta_b + 2\frac{\Gamma^{\mathrm{SM}}_d}{\Gamma^{\mathrm{SM}}_{\mathrm{had}}}\delta_d, \\ 
\delta_\sigma &=& \delta_{\mathrm{had}} + \delta_{l} -2 \delta_{Z}, \\ \label{corr3}
\delta_{A_f} &=& \frac{\delta g(f)_{1V}}{g(f)^{\mathrm{SM}}_{1V}} + \frac{\delta g(f)_{1A}}{g(f)^{\mathrm{SM}}_{1A}} - \delta_f,
\end{eqnarray}
where, for $f \neq b$
\begin{equation}
\delta_f = 2\frac{g(f)^{\mathrm{SM}}_{1V}\delta g(f)_{1V} + g(f)^{\mathrm{SM}}_{1A}\delta g(f)_{1A}}{(g(f)^{\mathrm{SM}}_{1V})^2 + (g(f)^{\mathrm{SM}}_{1A})^2},
\end{equation}
and for the bottom quark
\begin{equation}
\delta_b = \frac{(3-\beta^2)g(b)^{\mathrm{SM}}_{1V}\delta g(b)_{1V} + 2\beta^2 g(b)^{\mathrm{SM}}_{1A}\delta g(b)_{1A}}{\frac{3-\beta^2}{2}(g(b)^{\mathrm{SM}}_{1V})^2 + \beta^2(g(b)^{\mathrm{SM}}_{1A})^2}.
\end{equation}

The tree-level contribution to the $Z_1$ partial decays due to the $Z-Z^\prime$ mixing is included by multiplying $\Gamma(Z^{\mu}_1\rightarrow f\bar{f})$ in (\ref{decz}) by the factor $1 + \rho_V$, where $\rho_V \approx (M_{Z_2}^2/M_{Z_1}^2-1)\sin^2\theta$.

The theoretical value for the effective weak charge for the Cesium atom is given by \cite{bouchiat}
\begin{equation}\label{cesium}
Q_W(\mathrm{Cs})= Q^{\mathrm{SM}}_W(\mathrm{Cs})+ \Delta Q_W = Q^{\mathrm{SM}}_W(\mathrm{Cs}) \Bigl[ 1+ \delta_{Q_W}\Bigr],
\end{equation}
where \cite{durkin,altarelli}

\begin{equation}\label{deltaq}
\Delta Q_W=\left[Z \left(1+4\frac{S^4_W}{1-2S^2_W}\right)-N \right]\rho_V
+\Delta Q^\prime_W.
\end{equation}
with
\begin{eqnarray}\nonumber
\Delta Q^\prime_W &=& 16\lbrack (2Z+N)(g(e)_{1A}g(u)_{2V}+g(e)_{2A}g(u)_{1V}) \\ \nonumber
&+&(Z+2N)(g(e)_{1A}g(d)_{2V}+g(e)_{2A}g(d)_{1V})\rbrack \sin\theta \\ \nonumber
&-& 16\lbrack(2Z+N)g(e)_{2A}g(u)_{2V} \\
& & +(Z+2N)g(e)_{2A}g(d)_{2V}\rbrack
\frac{M^2_{Z_1}}{M^2_{Z_2}}. \label{deltaqp}
\end{eqnarray}
$Z$ and $N$ are, respectively, the number of protons and of neutrons in the nucleus of the considered atom. For the Cesium: $Z=55$ and $N=78$.

Clearly, $\Delta Q_W$ accounts for the contribution of the new physics. Notice that $\Delta Q^\prime_W$ is model dependent; in particular, it is a function of the couplings $g(q)_{2V}$ and $g(q)_{2A}$ ($q=u,d$) of the first family of quarks to the new neutral gauge boson $Z_2$. Because of this, the new physics in $\Delta Q^\prime_W$ depends on which family of quarks transforms differently under the gauge group. 

For the partial decays in (\ref{corr1}) we use \cite{pdg}
\begin{eqnarray}\nonumber
\Gamma^{\mathrm{SM}}_u = 300.10 \pm 0.09 \;\; \mbox{MeV}, \quad & \Gamma^{\mathrm{SM}}_\nu = 167.18 \pm 0.02 \;\; \mbox{MeV}, \\ \nonumber
\Gamma^{\mathrm{SM}}_d = 382.89 \pm 0.08 \;\; \mbox{MeV}, \quad & \Gamma^{\mathrm{SM}}_e = 83.97 \pm 0.03 \;\; \mbox{MeV}, \\ \label{Gammas}
\Gamma^{\mathrm{SM}}_b = 376.01 \pm 0.05 \;\; \mbox{MeV}. &  
\end{eqnarray}
\noindent

With the 3-4-1 predictions written in the form (\ref{corr}), we need the following well measured input parameters \cite{pdg}: $G_F = 1.166367(5)\times 10^{-5}$ GeV, $M_{Z_{1}}=91.1874\pm 0.0021$ GeV and $m_{t}=170.9 \pm 1.9$ GeV. For $S_W$ we use the value $\sin^2\widehat{\theta}_W(M_{Z_1})\equiv \widehat{s}^2_Z=0.23119 \pm 0.00014$ in the $\overline{\mbox{MS}}$ scheme because is less sensitive to $m_t$ than its value in the on-shell scheme, and for the bottom quark mass we use the running mass in the $\overline{\mbox{MS}}$ scheme at the $Z_1$ scale: $\widehat{m}_b(M_{Z_1})=2.67 \pm 0.19$ GeV \cite{Abbi}.

By using $g(e)_{iA}$ and $g(q)_{iV}, i=1,2$ from Tables \ref{tab:2} and \ref{tab:3}, and taking the third generation as the one transforming differently under $SU(4)_L \otimes U(1)_X$, the value we obtain for $\Delta Q_{W}^{\prime}$ is
\begin{equation}\label{dqpvalue}
\Delta Q_{W}^{\prime}=399.51\sin \theta -96.65\frac{M_{Z_{1}}^{2}}{M_{Z_{2}}^{2}}.
\end{equation}

\begin{table}
\caption{3-4-1 model predictions for the observables in Table \ref{tab:4}. The third column shows the percentage change in these observables relative to their SM values.}
\begin{indented}
\label{tab:5}
\item[]\begin{tabular}{lcc}
\br
3-4-1 model & Value & Percentage change\\
\mr
$\Gamma^{\mathrm{SM}}_Z[1+\delta_Z(1+\rho_V)]$\; [\mbox{GeV}] & $2.4979 \pm 0.0010$ & $0.045$ \\
$\Gamma^{\mathrm{SM}}(\mathrm{had})[1+\delta_{\mathrm{had}}(1+\rho_V)]$\; [\mbox{GeV}] & $1.7440 \pm 0.0013$ & $0.034$ \\
$\Gamma^{\mathrm{SM}}_{(l^{+}l^{-})}[1+\delta_l (1+\rho_V)]$\; [\mbox{MeV}] & $84.037 \pm 0.016$ & $0.058$ \\
$\sigma^{\mathrm{SM}}_{\mathrm{had}}(1 + \delta_\sigma)\; \lbrack \mbox{nb} \rbrack $ & $41.467 \pm 0.0217$ & $0.003$ \\
$R^{\mathrm{SM}}_e (1+\delta_{\mathrm{had}}-\delta_e)$ & $20.753 \pm 0.015$ & $-0.024$ \\
$R^{\mathrm{SM}}_{\mu} (1+\delta_{\mathrm{had}}-\delta_{\mu})$ & $20.753 \pm 0.015$ & $-0.024$ \\
$R^{\mathrm{SM}}_{\tau} (1+\delta_{\mathrm{had}}-\delta_{\tau})$ & $20.798 \pm 0.015$ & $-0.024$ \\
$R^{\mathrm{SM}}_b (1+\delta_b-\delta_{\mathrm{had}})$ & $0.21585 \pm 0.00012$ & $0.006$ \\
$R^{\mathrm{SM}}_c (1+\delta_c-\delta_{\mathrm{had}})$ & $0.17226 \pm 0.00009$ & $0.010$ \\
$A_{FB}^{(0,e)\mathrm{SM}}(1+2\delta_{A_e})$ & $0.01577 \pm 0.00022$ & $-3.05$\\
$A_{FB}^{(0,\mu)\mathrm{SM}}(1+\delta_{A_\mu}+\delta_{A_e})$ & $0.01577 \pm 0.00022$ & $-3.05$ \\
$A_{FB}^{(0,\tau)\mathrm{SM}}(1+\delta_{A_\tau}+\delta_{A_e})$ & $0.01577 \pm 0.00022$ & $-3.05$ \\
$A_{FB}^{(0,b)\mathrm{SM}}(1+\delta_{A_b}+\delta_{A_e})$ & $0.1017 \pm 0.0007$ & $-1.54$ \\
$A_{FB}^{(0,c)\mathrm{SM}}(1+\delta_{A_c}+\delta_{A_e})$ & $0.0726 \pm 0.0006$ & $-1.67$ \\
$A_{FB}^{(0,s)\mathrm{SM}}(1+\delta_{A_s}+\delta_{A_e})$ & $0.1018 \pm 0.0007$ & $-1.54$ \\
$A^{\mathrm{SM}}_{e}(1+\delta_{A_e})$ & $0.1450 \pm 0.0011$ & $-1.52$ \\
$A^{\mathrm{SM}}_{\mu}(1+\delta_{A_\mu})$ & $0.1450 \pm 0.0011$ & $-1.52$ \\
$A^{\mathrm{SM}}_{\tau}(1+\delta_{A_\tau})$ & $0.1450 \pm 0.0011$ & $-1.52$ \\
$A^{\mathrm{SM}}_{b}(1+\delta_{A_b})$ & $0.9345 \pm 0.0001$ & $-0.02$ \\
$A^{\mathrm{SM}}_{c}(1+\delta_{A_c})$ & $0.6668 \pm 0.0005$ & $-0.15$ \\
$A^{\mathrm{SM}}_{s}(1+\delta_{A_s})$ & $0.9534 \pm 0.0001$ & $-0.02$ \\
$Q^{\mathrm{SM}}_W(\mathrm{Cs}) [1+ \delta_{Q_W}]$ & $-72.71 \pm 0.03$ & $-0.61$\\
\br
\end{tabular}
\end{indented}
\end{table}

\begin{figure*} 
\begin{center}
\resizebox{0.98\textwidth}{!}{
\includegraphics{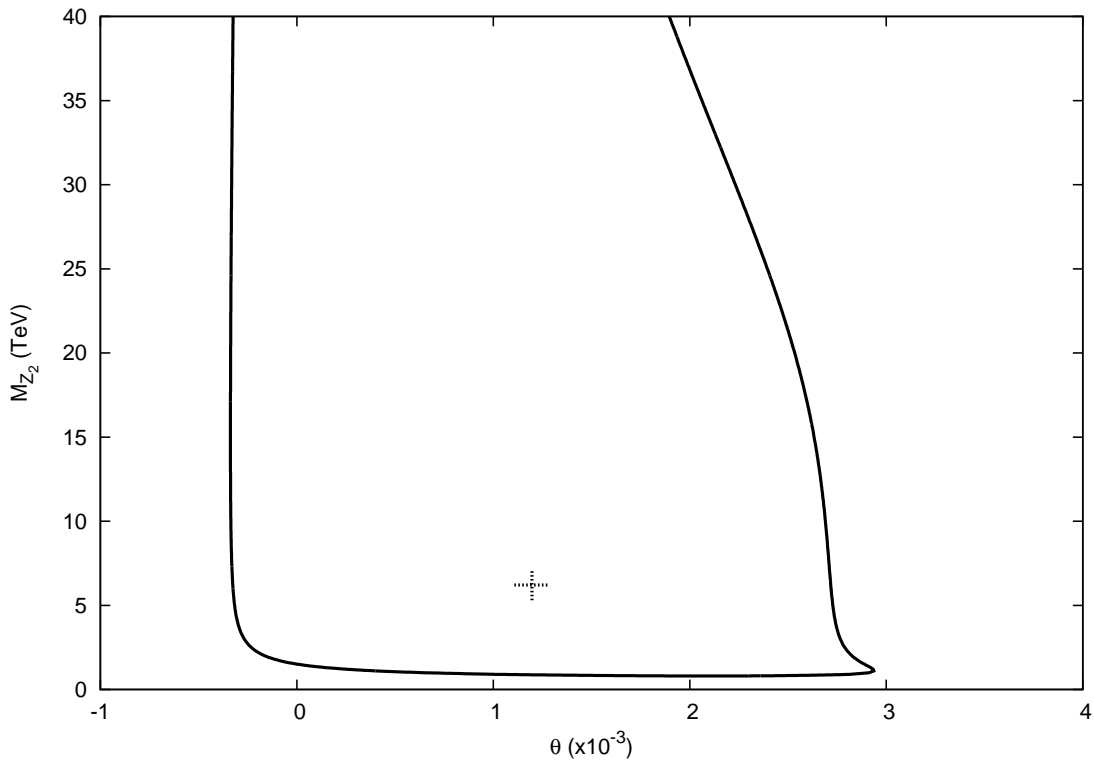}
}
\caption{\label{fig1} Contour plot displaying the allowed region for
$\theta$ vs $M_{Z_2}$ at $95\%$ C.L. from LEP, SLAC Linear Collider and APV data. The cross locates the best fit values}
\end{center}
\end{figure*}

Using the experimental values for the $Z$-pole observables in Table \ref{tab:4} and with $\Delta Q_W$ in terms of new physics in (\ref{deltaq}), we do a $\chi^2$ fit of the theoretical expressions in Table \ref{tab:5} to the data and find the best allowed region in the $(\theta - M_{Z_2})$ plane at $95\%$ confidence level (C.L.). This region is shown in Figure \ref{fig1} which provide us the constraints
\begin{equation}\label{bounds}
-0.00034 \leq\theta\leq 0.00294, \quad 0.802\; {\mbox TeV} \leq M_{Z_2}.
\end{equation}
The fit has a $\chi^2/\mathrm{d.o.f.}$ of $17.7/20$, corresponding to a probability of $60.7\%$, and the best-fit values are: $\theta = 0.00120$, $M_{Z_2}=6.205$ TeV.

We have evaluated the effect of the uncertainty in $\widehat{s}^2_Z$ on the constraints in (\ref{bounds}), because this is the most correlated input parameter. To this purpose we have left $\widehat{s}^2_Z$ free to vary in the fit subject to the constraint $\widehat{s}^2_Z = 0.23119 \pm 0.00014$. In this case the bounds change to

\begin{equation}\label{bounds2}
-0.00064 \leq\theta\leq 0.00316, \quad 0.780\; {\mbox TeV} \leq M_{Z_2},
\end{equation}
where the second one represents a percentage change, relative to the limit in (\ref{bounds}), of only $2.74\%$ in the lower bound on $M_{Z_2}$.
  
As we can see, the lower bounds on $M_{Z_2}$ are compatible with the bound obtained in $p\bar{p}$ collisions at the Fermilab Tevatron \cite{abe}.

Here we point out that the bounds in (\ref{bounds}) correct the bounds $-0.0032\leq \theta \leq 0.0031$ and $0.67\; \mathrm{TeV}\leq M_{Z_2}\leq 6.1\; \mathrm{TeV}$ obtained in \cite{spp}. The latter are incorrect because do not satisfy $M_{Z_2} \rightarrow \infty$ in the limit $\theta \rightarrow 0$, that is, an upper bound of $6.1$ TeV must not occur.

Using the best-fit values for $\theta$ and $M_{Z_2}$, we calculate the 3-4-1 model predictions for the electroweak observables in Table \ref{tab:4} and the percentage changes in these observables relative to their SM values. As a first approximation we neglect correlations between the uncertainties of the input parameters and use standard error propagation. This is partially justified because the errors which enter in the expressions for the 3-4-1 predictions are of different status and, therefore, there is no clean way of exactly calculating the errors. The results are shown in Table \ref{tab:5}. Notice that, except for the forward-backward asymmetries and the obsevables $A_e$, $A_\mu$ and $A_\tau$, the other percentage changes are at the per-mille level and even lower. In any case, no substantial improvement to the SM fit is observed.

We remark that, since in this model only $SU(2)_L$ scalar singlets and doublets develop VEV, the $Z-Z^\prime$ mixing contribution to the $\rho$ parameter is such that $\rho_V << 1$. This, together with the fact that all the 3-4-1 model corrections $\delta O$ to SM observables go to zero in the limit $\theta \rightarrow 0$ ($M_{Z^\prime} \rightarrow \infty$), justifies our fitting procedure in which we treat the new physics effects as small corrections to the well established SM results \cite{Cza-Cha}.

As already mentioned, the bounds in (\ref{bounds}) are obtained assuming that is the third generation of quarks the one transforming differently under $SU(4)_L\otimes U(1)_X$. Notice however from Table \ref{tab:1} that, as remarked above, the three quark families have the same hypercharge $X$ with respect to the $U(1)_X$ subgroup. As a result, the couplings $g(f)_{iA}$ and $g(f)_{iV}, i=1,2$ of all the fermion fields to $Z_1$ and $Z_2$ are family universal, which in turn implies that for this model the constraints in (\ref{bounds}) do not depend on which family of quarks transforms differently.

\subsection{Bounds on $M_{Z_3}$ from FCNC processes}
\label{sec:42}

After eliminating the source of FCNC associated to the mixing between ordinary and exotic fermions by the introduction of the discrete $Z_2$ symmetry, there remains two sources of FCNC in the model. The first one is identified by noticing that, since each flavour couples to more than one Higgs 4-plet, there are FCNC coming from the scalar sector. Because this contribution depends on the large number of arbitrary parameters in the scalar potential, is not very useful to constrain the model and we will ignore it. The second source, which is also the only one if we neglect the scalar contribution, comes from the left-handed interactions of ordinary quarks with the neutral gauge boson $Z^{\prime \prime}$ which, as we already know, are flavor nondiagonal. For their study we will follow the analysis presented in \cite{liu,rodriguez} where bounds coming from neutral meson mixing are obtained in the framework of the so-called ``minimal 3-3-1 model".

From the charge generator in (\ref{carga}), the value of the $Y$ hypercharge of the SM is obtained as: $Y/2=T_{8L}/\sqrt{3}-2T_{15L}/\sqrt{6}+X$. Using this expression, the couplings of $Z^{\prime\prime}$ to left-handed quarks in (\ref{zprimaprima}), can be written in a more convenient fashion for $4$-plets as
\begin{equation}\label{lagzprimaprima}
\mathcal{L}(Z^{\prime \prime })=-\frac{g}{\sqrt{3}}Z_{\mu }^{\prime \prime
}J^{\mu}(Z^{\prime \prime })= -\frac{g}{\sqrt{3}}Z_{\mu }^{\prime \prime
}\sum_{f}\bar{f}\gamma ^{\mu}T_{L}P_{L}f,
\end{equation}
where $P_{L}$ is the left-handed projection operator and $T_{L}=\sqrt{2}T_{8L}+T_{15L}$.

Since the value of the operator $T_L$ is different for $4$-plets than for $4^*$-plets, the flavour changing interaction can be written, for ordinary up- and down-type quarks $q^\prime$ in the weak basis, as
\begin{equation}\label{jfcnc}
J^{\mu }(Z^{\prime \prime })_{\mathrm{FCNC}}=\sum_{q^{\prime }}\bar{
q^{\prime }}\gamma ^{\mu }[T_{L}(4)-T_{L}(4^{\ast })]P_{L}q^{\prime }.
\end{equation}
From (\ref{lagzprimaprima}) and (\ref{jfcnc}) we get
\begin{equation}\label{fcnczprimaprima}
\mathcal{L}(Z^{\prime \prime })_{\mathrm{FCNC}}=-\frac{g}{\sqrt{2}}Z_{3}^{\mu
}\sum_{q^{\prime }}\bar{q^{\prime }}\gamma _{\mu
}P_{L}q^{\prime }.
\end{equation}

Using (\ref{fcnczprimaprima}) we will deduce constraints on the $Z_3$ mass coming from experimental data in the $K^0-\bar{K}^0$, $B^0_d-\bar{B}^0_d$, $B^0_s-\bar{B}^0_s$ and $D^0-\bar{D}^0$ systems. To this purpose we recall that the mass matrices $M^u$ and $M^d$ in (\ref{mu}) and (\ref{md}) are diagonalized by biunitary transformations $U_{L,R}$ and $V_{L,R}$, respectively, with $V_{CKM}= U^{\dagger}_LV_L$ being the CKM mixing matrix. Then, in terms of mass eigenstates, (\ref{fcnczprimaprima}) produces the following effective Hamiltonian for the tree-level neutral meson mixing interactions
\begin{equation}\label{heff}
\mathcal{H}_{\mathrm{eff}}^{(\alpha ,\beta )}=\sqrt{2}G_{F}C_{W}^{2}\left(
V_{Lj\alpha }^{\ast }V_{Lj\beta }\right) ^{2}\frac{M_{Z_{1}}^{2}}{
M_{Z_{3}}^{2}}\left( \bar{\alpha }\gamma _{\mu }P_{L}\beta \right) ^{2},
\end{equation}
where $(\alpha,\beta)$ must be replaced by $(d,s)$, $(d,b)$, $(s,b)$ and $(u,c)$ for the $K^0-\bar{K}^0$, $B^0_d-\bar{B}^0_d$, $B^0_s-\bar{B}^0_s$ and $D^0-\bar{D}^0$ systems, respectively, and $V_L$ must be replaced by $U_L$ for the neutral $D^0-\bar{D}^0$ system. The family index $j= 1,2,3$ refers to the family of quarks to be singled out as transforming differently under $SU(4)_L$.

If we assume that the heaviest family of quarks is the one transforming differently, the effective Hamiltonian gives the following contribution to the mass difference $\Delta m_K$
\begin{equation}\label{deltamk}
\frac{\Delta m_{K}}{m_{K}}=\frac{2\sqrt{2}G_{F}C_{W}^{2}}{3}Re[\left(
V_{L3d}^{\ast }V_{L3s}\right) ^{2}]\eta _{K}\frac{M_{Z_{1}}^{2}}{
M_{Z_{3}}^{2}}B_{K}f_{K}^{2},
\end{equation}
while for the $B^0_d-\bar{B}^0_d$, $B^0_s-\bar{B}^0_s$ and $D^0-\bar{D}^0$ systems, we have
\begin{equation}\label{deltamb}
\frac{\Delta m_{B}}{m_{B}}=\frac{2\sqrt{2}G_{F}C_{W}^{2}}{3}\vert
V_{L3\alpha }^{\ast}V_{L3\beta }\vert ^{2}\eta _{B}\frac{M_{Z_{1}}^{2}}{
M_{Z_{3}}^{2}}B_{B}f_{B}^{2},
\end{equation}
\begin{equation}\label{deltamd}
\frac{\Delta m_{D}}{m_{D}}=\frac{2\sqrt{2}G_{F}C_{W}^{2}}{3}\vert
V_{L3u}^{\ast }V_{L3c}\vert ^{2}\eta _{D}\frac{M_{Z_{1}}^{2}}{
M_{Z_{3}}^{2}}B_{D}f_{D}^{2},
\end{equation}
where the subindex $B$ in (\ref{deltamb}) stands for $B_d$ or $B_s$. $B_m$ and $f_m$ ($m=K, B_d, B_s,$ $D$) are the bag parameter and decay constant of the corresponding neutral meson. The $\eta$'s are QCD correction factors which, at leading order, can be taken equal to the ones of the SM \cite{blanke}, that is: $\eta_K\simeq \eta_D\simeq 0.57$, $\eta_{B_d} = \eta_{B_s}\simeq 0.55$ \cite{gilman}.

\begin{table}
\caption{Values of the experimental and theoretical quantities used as input parameters for FCNC processes.}
\begin{indented}
\label{tab:6}
\item[]\begin{tabular}{lcc}
\br
& Value & Reference \\
\mr
$\Delta m_K$\; [\mbox{GeV}] & $3.483(6)\times 10^{-15}$ & \cite{pdg} \\
$m_{K^0}$\; [\mbox{MeV}] & $497.65(2)$ & \cite{pdg} \\
$f_K\sqrt{B_K}$\; [\mbox{MeV}] & $143(7)$ & \cite{hashimoto} \\ \mr
$\Delta m_{B_d}$\; [$\mbox{ps}^{-1}$] & $0.508(4)$ & \cite{pdg} \\
$m_{B_d}$\; [\mbox{GeV}] & $5.2794(5)$ & \cite{pdg} \\
$f_{B_d}\sqrt{B_{B_d}}$\; [\mbox{MeV}] & $214(38)$ & \cite{hashimoto} \\ \mr
$\Delta m_{B_s}$\; [$\mbox{ps}^{-1}$] & $17.77(12)$ & \cite{abulencia} \\
$m_{B_s}$\; [\mbox{GeV}] & $5.370(2)$ & \cite{pdg} \\
$f_{B_s}\sqrt{B_{B_s}}$\; [\mbox{MeV}] & $262(35)$ & \cite{hashimoto} \\ \mr
$\Delta m_D$\; [$\mbox{ps}^{-1}$] & $11.7(6.8)\times 10^{-3}$ & \cite{ciuchini} \\
$m_{D^0}$\; [\mbox{GeV}] & $1.8645(4)$ & \cite{pdg} \\
$f_D\sqrt{B_D}$\; [\mbox{MeV}] & $241(24)$ & \cite{artuso} \\ \mr
$m_u(M_Z)$\; [\mbox{MeV}] & $2.33^{+0.42}_{-0.45}$ & \cite{fusaoka} \\
$m_c(M_Z)$\; [\mbox{MeV}] & $677^{+56}_{-61}$ & $$ \\
$m_t(M_Z)$\; [\mbox{GeV}] & $181\pm 13$ & $$ \\
$m_d(M_Z)$\; [\mbox{MeV}] & $4.69^{+0.60}_{-0.66}$ & $$ \\
$m_s(M_Z)$\; [\mbox{GeV}] & $93.4^{+11.8}_{-13.0}$ & $$ \\
$m_b(M_Z)$\; [\mbox{GeV}] & $3.00\pm 0.11$ & $$ \\
\br
\end{tabular}
\end{indented}
\end{table}

In order to obtain limits on $M_{Z_3}$ from the former equations, two remarks are in order: (1) it is well known that the complex numbers $V_{Lij}$ and $U_{Lij}$ cannot be estimated from the present experimental data. We overcome this obstacle by assuming the Fritzsch ansatz for the quark mixing matrix \cite{fritzsch}, which implies (for $i\leq j$) $V_{Lij}=\sqrt{m_i/m_j}$, and similarly for $U_L$ \cite{cheng} (CP violating phases in the mixing matrices will not be considered here); (2) the contribution of the $Z_3$ exchange to the mass differences is not the only one. Several sources may also contribute to them and it is not possible to disentangle the $Z_2$ contribution from other effects. Because of this, several authors consider reasonable to assume that the $Z_2$ exchange contribution must not be larger than the experimental values \cite{liu}. In this work we will assume that this is the case. We must notice, however, that more conservative but rather arbitrary criteria have been used by other authors \cite{blanke}.

With these ingredients, we obtain bounds on $M_{Z_3}$ by using updated experimental and theoretical values for the input parameters as shown in Table~\ref{tab:6}. For each neutral meson system, the results are
\begin{eqnarray}\nonumber
K^0-\bar{K}^0: & M_{Z_3}>2.40\; \mbox{TeV}, \\ \nonumber
B^0_d-\bar{B}^0_d: & M_{Z_3}>6.65 \;\mbox{TeV}, \\ \nonumber
B^0_s-\bar{B}^0_s: & M_{Z_3}>6.19\; \mbox{TeV}, \\ \label{fcnc}
D^0-\bar{D}^0: & M_{Z_3}>0.16\; \mbox{TeV}.
\end{eqnarray}

This shows that the strongest constraint comes from the $B^0_d-\bar{B}^0_d$ system, which poses a lower bound on $M_{Z_3}$ larger than $6.65\; \mbox{TeV}$.

A detailed analysis shows that the constraints in (\ref{fcnc}) are family dependent in the sense that their values change according the family of quarks chosen as the one transforming differently under $SU(4)_L$. When the first or the second family are chosen, the strongest constraint on the lower bound on $M_{Z_3}$ comes from the $K^0-\bar{K}^0$ system and turns out to be larger than $\sim 75$ TeV \cite{NS}.
 
\section{Summary and conclusions}
\label{sec:5}

In this work, in the context of the extension of the SM based on the gauge group $SU(3)_C\otimes SU(4)_L\otimes U(1)_X$, which predict the existence of two extra neutral gauge bosons $Z^\prime $ and $Z^{\prime\prime}$, we have set bounds on the mixing angle $\theta$ between the SM gauge boson $Z$ and the new $Z^\prime$, and on the masses of the new physical eigenstates, namely: $Z_2$, which arises from the diagonalization of the $Z-Z^\prime$ mass matrix, and $Z^{\prime\prime} \equiv Z_3$ which becomes a mass eigenstate with $M^2_{Z_3}=(g^2_4/2)(V^2$ $+v^2)$ when the conditions $V^\prime \simeq V$ and $v^\prime \simeq v$ are fulfilled. $V^\prime$, $V$, $v^\prime$ and $v$ are the vacuum expectation values of four Higgs $4^*$-plets used to break the symmetry. We have assumed the hierarchy $V^\prime \simeq V >> v^\prime \simeq v$, and from the mass of the lightest charged gauge boson $M_{W^{\pm}}^{2}=(g_{4}^{2}/2)(v^{2}+v^{\prime 2})$, that can be identified with the SM $W$ boson, we have obtained $\sqrt{v^{\prime 2}+v^2}\simeq v_{\mathrm{EW}}=174$ GeV. The other charged gauge bosons in the model acquire masses at the large scale $V^\prime \simeq V$.

We have studied a version of the 3-4-1 extension characterized by the values $b=1,\; c=-2$ of the parameters appearing in the electric charge operator in (\ref{carga}), with fermion content without exotic electric charges and with anomalies cancelling among the fermion families in a non-trivial fashion. This method of cancellation of anomalies leads to a number of fermion families $N_f$ that must be divisible by the number of colours $N_c$ of $SU(3)_c$, being $N_f = N_c = 3$ the simplest solution. In this last case universality in the lepton sector is preserved, but one family of quarks must transform differently than the other two under $SU(4)_L\otimes U(1)_X$. This fact leads to FCNC arising at the tree-level and transmitted, in the model studied here, by the neutral gauge bosons $Z_3$.

The limits on the parameters $\theta$ and $M_{Z_2}$ have been obtained by doing a $\chi^2$ fit of the theoretical predictions of the 3-4-1 model, for 22 precision electroweak observables, to the experimental data at the Z-pole from LEP and SLAC Linear Collider and atomic parity violation data. We have obtained: $-0.00034 \leq\theta\leq 0.00294$ and $M_{Z_2} \geq 802$ GeV. These bounds correct the ones reported in \cite{spp}. The mass of the additional new neutral gauge boson $Z_3$ has been constrained by using experimental data from neutral meson mixing in the study of the FCNC effects associated to quark family nonuniversality. For the calculation we have assumed the Fritzsch ansatz for the quark mixing matrices and we have taken complex phases equal to zero. In this way we have found that the strongest constraint comes from the $B^0_d-\bar{B}^0_d$ system, which poses a lower bound on $M_{Z_3}$ larger than $6.65$ TeV. It must be however recognized that the bounds from neutral meson mixing are obscured by the lack of knowledge of the entries in the quark mixing matrices and by the rather arbitrary assumed contribution of the $Z_3$ exchange to the mass differences in the neutral meson systems.

We also have done a comparison between the predictions of the 3-4-1 model studied here and the predictions of the SM for the 22 observables mentioned above. We have found that the 3-4-1 model fits the data at least as well as the SM does.

We have forbidden mixing between ordinary and exotic fermions, which in turn avoids violation of the unitarity of the CKM mixing matrix, by introducing an anomaly free discrete $Z_2$ symmetry under which the SM particles are singlets. This symmetry, combined with the four Higgs scalars, generates a consistent charged fermion mass spectrum with the ordinary charged fermions acquiring masses at the low scale $v^\prime \simeq v$ and with the exotic charged fermions getting masses at the high scale $V^\prime \simeq V$. The neutral leptons remain massless after the symmetry breaking. Notwithstanding, the extension of the original scalar sector in (\ref{vev}) and the inclusion of neutral fermions, singlets under $SU(4)_L$ and with zero $X$ charges, can accomodate neutrino phenomenology in the model \cite{riazu}.

3-4-1 models in the $b=1,\; c=-2$ class, as the one studied in this paper, have the particular feature that, notwithstanding two families of quarks transform differently under the $SU(4)_L$ subgroup, the three families have the same hypercharge $X$ with respect to the $U(1)_X$ subgroup. As a consequence, the couplings of the ordinary fermion fields to the neutral currents $Z_1$ and $Z_2$ are family universal. Thus, the allowed region in the parameter space $\theta-M_{Z_2}$ and the lower limit $M_{Z_2} \geq 0.802$ TeV do not depend on which family of quarks transforms differently under the gauge group. Since FCNC are present for this Model in the left-handed couplings of ordinary quarks to the $Z_3$ gauge boson, the contribution of the $Z_3$ exchange to the mass differences in neutral meson systems produces family-dependent constraints on the $Z_3$ mass. The detailed study in \cite{NS} shows that the third family of quarks must transform differently in order to get the smallest lower bound on $M_{Z_3}$ which, as said above, comes from the $B^0_d-\bar{B}^0_d$ system and turns out to be $M_{Z_3}>6.65$ TeV. Since $M^2_{Z_3}=(g^2_4/2)(V^2$ $+v^2)$, this is also a lower bound on the scale of breaking of the 3-4-1 symmetry. So, the heaviest family of quarks must be the one transforming differently if we want to end with a 3-4-1 scale of the order of a few TeV and, consequently, with a model able to be tested at the LHC facility.

By contrast, for 3-4-1 models in the $b=c=1$ class, anomaly cancellation between generations implies not only a family of quarks transforming differently than the other two, but also a nonuniversal hypercharge $X$ for the left-handed quark multiplets (for details see \cite{swz,sgp}). So, the couplings $g(f)_{iV}$ and $g(f)_{iA}$ ($i=1,2$) of ordinary fermions to the neutral currents $Z_1$ and $Z_2$ are family dependent, which implies that in this case the allowed region in the parameter space $\theta-M_{Z_2}$ depends on which family of quarks transforms differently under $SU(4)_L\otimes U(1)_X$. The analysis leads to the conclusion stated in \cite{NS} according to which, also in this class of models, the third family of quarks must transform differently in order to have a lower bound on $M_{Z_2}$ as low as possible, which turns out to be 2 TeV. Moreover, in this class of models the left-handed couplings of $Z_2$ to the SM quarks are flavour nondiagonal which induces tree level FCNC transmitted by this extra neutral gauge boson. As shown in \cite{swz}, the constraints coming from FCNC data are also family-dependent and, provided the heaviest family of quarks transforms differently, they raise the lower limit on $M_{Z_2}$ obtained from the fit to Z-pole observables (2 TeV), to a value larger than $\sim 12$ TeV. Bounds on the mass of $Z^{\prime\prime}\equiv Z_3$ are not obtained because this current couples only to exotic fermions and thus decouples completely from the low energy physics.
 
The former considerations allows us to conclude that the $b=1,\;c=-2$ class of 3-4-1 models are favoured in the sense that they provide the smallest lower bounds on $M_{Z_2}$ and $M_{Z_3}$ which are in the range $(1-10)$ TeV and, consequently, they have a better chance to be tested at the LHC or further at the ILC. 

The particular conditions under which new heavy resonance peaks possibly occurring at the LHC (for example in Drell-Yan dilepton production) could be identified as having their origin in the class of 3-4-1 models studied here, deserve attention and will be discussed elsewhere.

\ack
We acknowledge financial support from DIME at Universidad Nacional de Colombia-sede Medell\'\i ­n and from COLCIENCIAS in Colombia under contract 1115-333-18740. We also thank Alejandro Jaramillo for helping us to refine the computer program used for the numerical analysis presented in section \ref{sec:4}.

\end{document}